\documentclass[aps,prd,twocolumn,superscriptaddress,amsfont,amssymb,amsmath,nofootinbib,showpacs,balancelastpage]{revtex4-1}
	\usepackage[english]{babel}
	\usepackage{amsmath,amssymb,amsxtra}					
	
	\usepackage{graphicx}
\usepackage[latin1]{inputenc}
\usepackage{amsmath}
\usepackage{amsfonts}
\usepackage{appendix}
\usepackage{amssymb}
\usepackage{epsfig}
\usepackage{graphicx}
\usepackage{dcolumn}
\usepackage{bm}
\usepackage{epstopdf}
\usepackage{color}
\usepackage[usenames,dvipsnames,svgnames]{xcolor}
\usepackage[colorlinks=true,
            linkcolor=red,
            urlcolor=gray,
            citecolor=blue]{hyperref}
\usepackage{multirow}
\usepackage{float}

\usepackage{subfigure}

\usepackage{enumitem}

\def\3nab{\tilde{\nabla}}

\def\be {\begin{equation}}
\def\ee {\end{equation}}
\def\ba {\begin{align}}
\def\ea {\end{align}}

\def\bc {\begin{center}}
\def\ec {\end{center}}
\def\case#1/#2{\frac{#1}{#2}}

\newcommand{\bver}{\begin{verbatim}}
\newcommand{\ever}{\end{verbatim}}

\newcommand{\bea}{\begin{eqnarray}}
\newcommand{\eea}{\end{eqnarray}}
\newcommand{\beaa}{\begin{eqnarray*}}
\newcommand{\eeaa}{\end{eqnarray*}}

\def\case#1/#2{\textstyle\frac{#1}{#2}}

\newcommand{\lsim}   {\mathrel{\mathop{\kern 0pt \rlap
  {\raise.2ex\hbox{$<$}}}
  \lower.9ex\hbox{\kern-.190em $\sim$}}}
\newcommand{\gsim}   {\mathrel{\mathop{\kern 0pt \rlap
  {\raise.2ex\hbox{$>$}}}
  \lower.9ex\hbox{\kern-.190em $\sim$}}}

\begin{document}
\title{On the tension between growth rate and CMB data}

\author{Alba Quelle}
\email{aquelle@ucm.es}
\author{Antonio L. Maroto}
\email{maroto@ucm.es}
\affiliation{Departamento de F\'{\i}sica Te\'orica and Instituto de F\'isica de 
Part\'iculas y del Cosmos (IPARCOS), Universidad Complutense de Madrid, 28040 
Madrid, Spain}

\pacs{98.80.-k}



\begin{abstract} 
We analyze the claimed tension between redshift space distorsions measurements of $f(z)\sigma_8(z)$ and the predictions of 
standard $\Lambda$CDM (Planck 2015 and 2018) cosmology. We consider a dataset consisting of 17 data points extending up to redshift $z=1.52$ and corrected for the Alcock-Paczynski effect. Thus, calculating the evolution of the growth factor in a $w$CDM cosmology, we find that the tension for the best fit parameters $w$, $\Omega_m$ and $\sigma_8$ with respect to the Planck 2018 $\Lambda$CDM parameters is below $2\sigma$ in all the marginalized confidence regions. 
 \end{abstract} 


\maketitle
\section{Introduction}
Large-scale galaxy surveys are becoming one of the most powerful tools to test the
currently accepted  $\Lambda$CDM model based on General
Relativity. The possibility of mapping the distribution of matter in large
volumes at different redshifts allows to measure the growth rate of structures as a function 
of time and (length) scale which is a well-defined prediction of 
any cosmological model. 

The ability of such surveys to construct 3D maps depends crucially 
on the precise determination of galaxy redshifts from which  
radial distances to the survey objects can be inferred. The actual conversion depends, in turn, on
two important effects. On one hand, peculiar velocities introduce  distorsions in the 
redshift distribution, the so called redshift space distorsions (RSD), generating 
an anisotropic galaxy power spectrum. On the other, 
although at low redshift the Hubble law provides a straightforward relation between 
resdhift and distances, at higher redshifts this conversion depends on the chosen
fiducial cosmology. This fact lays behind the Alcock-Paczynski (AP) effect. In recent times these effects 
have allowed  to measure the linear growth rate of structures, defined 
as $f=d\ln\delta_m/d\ln a$ with $\delta_m$ the linear matter density contrast
with relatively good precision in a wide range of redshifts.
More precisely,  RSD provide a measurement
of the quantity $f(z)\sigma_8(z)$, where $\sigma_8(z)$ is the normalization
of the linear matter power spectrum at redshift $z$ on scales of $8 h^{-1}$Mpc.
In particular, measurements which can reach $10\%$ precision  have been obtained at $z<1$ by different surveys 
such as 2dF \cite{Percival:2004fs}, 6dFGRS \cite{Beutler:2012px}, WiggleZ \cite{Blake:2011rj} and recently by SDSS-III BOSS \cite{Alam:2016hwk} and
VIPERS \cite{Pezzotta:2016gbo}. At higher redshifts two measurements have also been
obtained recently by FMOS \cite{Okumura:2015lvp} and  from the BOSS quasar sample \cite{Zarrouk:2018vwy} although with relatively lower
precision.

Confrontation of $f(z)\sigma_8(z)$ measurements with  standard 
$\Lambda$CDM cosmology  predictions  has lead in recent years to claims of inconsistency or 
tension at different statistical significance levels. Thus in \cite{Macaulay:2013swa} a lower growth rate than expected from Planck $\Lambda$CDM cosmology 
was identified for the first time. Later on  \cite{MacCrann:2014wfa,Battye:2014qga} a tension at the 2$\sigma$ level
was claimed between the Planck data and the CFHTLenS determination
of $\sigma_8$. A similar tension was found by the KiDS+VIKING tomographic 
shear analysis \cite{Hildebrandt:2018yau}. More recently \cite{Nesseris:2017vor} a 3$\sigma$ tension with respect to the best fit parameters of Planck 2015 was also identified in a set of 18 data points
from RSD measurements of $f(z)\sigma_8(z)$. The tension could even increase up to $5\sigma$
if an extended dataset is used \cite{Kazantzidis:2018rnb}. The extended dataset has far more data at low redshifts where the model discrimination is easier \cite{Kazantzidis:2018jtb}, however in this extended case possible correlations within data points have not been taken into account\footnote{See however \cite{Skara:2019usd} for a more recent analysis}.   The possibility that more recent datapoints with larger errorbars compared to  earlier datapoints could introduce a bias  towards the expected standard Planck/$\Lambda$CDM cosmology is also discussed in \cite{Kazantzidis:2018rnb}. 

In this paper,  we follow a complementary approach, rather than using 
the extended dataset for which correlations are unknown, we will revisit the analysis of the tension from the most conservative point of view, i.e. using only independent datapoints or points whose correlations are known. Thus we consider the Gold dataset of  \cite{Nesseris:2017vor} and introduce
two changes in the analysis.  Firstly, we include
the most recent measurements from BOSS-Q \cite{Zarrouk:2018vwy} and 
on the other, we draw attention on the possible correlation between SDSS-LRG and BOSS-LOWZ
points.  Notice that the former are obtained from the SDSS data release DR7 with practically the 
same footprint as the later, obtained from DR10 and DR11,  but with less galaxies. 
In this sense, we explore the consequences of removing the two SDSS-LRG points from the analysis. In this sense, our dataset is very similar to that consider by Planck collaboration \cite{Aghanim:2018eyx}.
On the other hand, we will take the best fit parameters of Planck 2018 (CMB alone) given in  \cite{Aghanim:2018eyx} rather than Planck 2015 used in \cite{Nesseris:2017vor}. We consider the same type of $w$CDM cosmologies with three free parameters $(w,\Omega_m,\sigma_8)$
but obtain the confidence regions from the marginalized (rather than maximized) likelihoods. This enlarges the confidence regions so that 
the tension is found to be reduced below the $2\sigma$ level for all the parameters combinations.

\section{Growth of structures and $f(z)\sigma_8(z)$}
Let us consider a flat Robertson-Walker background whose metric in conformal time 
reads
\begin{eqnarray}
ds^2 = a^2 (\eta) \left[ -  d\eta ^2 + \delta _{ij} dx^i dx^j \right] 
\end{eqnarray}

The evolution of matter density perturbations $\delta_m=\delta\rho_m/\rho_m$ in a general cosmological model with 
non-clustering dark energy and standard conservation of matter is given for sub-Hubble scales by
\begin{equation}
	\delta _m '' + \mathcal{H} \delta _m ' - \frac{3}{2} \mathcal{H}^2 \Omega _m(a) \delta _m = 0 \label{deltam}
\end{equation} 
where prime denotes derivative with respect to conformal time, ${\cal H}=a'/a$
and $\Omega_m(a)=\rho_m/(\rho_m+\rho_{DE})$.  In this work we will limit ourselves to 
$w$CDM cosmologies so that at late times
\begin{eqnarray}
{\cal H}^2=H_0^2 a^2(\Omega_m a^{-3}+(1-\Omega_m)a^{-3(1+w)})
\end{eqnarray}
and 
\begin{eqnarray}
\Omega _m(a) = \frac{\Omega _{m}}{\Omega _{m} + (1-\Omega_m) \  a^{-3w}}
\end{eqnarray}
The growth rate is defined as
\begin{eqnarray}
f=\frac{d\ln\delta_m}{d\ln a}
\end{eqnarray}
which can be approximated by $f\simeq \Omega_m^\gamma(a)$ with $\gamma\simeq 0.55$ for $w$CDM models. Even 
though this fitting function provides accurate description for cosmologies close to $\Lambda$CDM, since we are interested in exploring a wide range of parameter space, in this work we will obtain $f$ just by numerically solving (\ref{deltam}) with initial conditions $\delta_m(a_i)=1$ and $\delta'(a_i)=1/a_i$ with $a_i$ well inside the matter dominated era.

The matter power spectrum corresponding to the matter density contrast in Fourier 
space $\delta_k(z)$ with $1+z=1/a$ is given by  $P(k,z)=V\vert \delta_k(z)\vert^2$ with $V$ the volume.
 Thus the variance of the matter 
fluctuations on a scale $R$ is given by
\begin{equation}
	\sigma _R ^2(z) = \frac{1}{2\pi ^2} \int P(k',z) W_R ^2 (k') ^2 dk'
\end{equation}
with the window function defined as:
\begin{eqnarray}
W_R(k)=\frac{3}{k^{3}R^{3}} \, [\sin(kR)-kR\cos(kR)]
\end{eqnarray}
Thus $\sigma_8(z)$ corresponds to $\sigma_R(z)$ at the scale $R= 8 h^{-1}$ Mpc.

From the matter power spectrum it is possible to define the galaxy power spectrum as
$P_g(k,z)=b^2(z)P(k,z)$ with $b(z)$ the bias factor.

From the observational point of view, galaxy surveys are able to determine the 
galaxy power spectrum in redshift space, which is given by
\begin{eqnarray}
&&P_{r,\textrm{obs}} (k_r,\mu _r ; z) \label{gal} \\
&=&\frac{H(z) d_{Ar} ^2 (z)}{H_r (z) d_A^2 (z)} D^2 (z) b^2 (z) \left[ 1 + \beta (z) \mu ^2
 \right] ^2 P(k,z=0) \nonumber
\end{eqnarray} 
where $H(z)=(1+z){\cal H}(z)$, 
\begin{eqnarray}
d_A (z) = \frac{1}{1+z} \int_{\frac{1}{1+z}}^1 \frac{1}{a^{1/2}} \frac{da }{H_0 \sqrt{\Omega _{m}  + (1-\Omega _{m} ) \  a^{-3w}}} 
\end{eqnarray}
is the angular diameter distance, $D(z)=\delta_m(z)/\delta_m(0)$ is the growth factor, $\beta(z) = f(z)/b(z)$  and $\mu$ is the cosine of the angle between $\hat k$ and 
the observation direction. Finally, the index $r$ denotes that the corresponding quantity is 
evaluated on the fiducial cosmology. Notice that the first factor in (\ref{gal}) 
corresponds to the AP effect, whereas the $(1+\beta\mu^2)^2$ factor is generated 
by the RSD. As we see RSD induce an angular dependence on the power spectrum which
contains a monopole, quadrupole and  hexadecapole contributions. From the measurements of  
monopole and quadropole it is possible to obtain the $f(z)\sigma_8(z)$ function that
for simplicity in the following we will simple denote $f\sigma_8(z)$. 
The measured value depends on the fiducial cosmology, so that in order to translate
from the fiducial cosmology used by the survey to other cosmology it 
is needed to rescale by a factor \cite{Nesseris:2017vor}
\begin{eqnarray}
\textrm{ratio}(z) = \frac{H(z) d_A (z)}{H_r(z) d_{A,r}(z)} \label{ratio}
\end{eqnarray} 
The fiducial cosmology correction could affect not only $f\sigma_8$ but also the power spectrum or even introduce additional multipoles in the galaxy power spectrum in redshift space. In principle, all these effects could be properly taken into account but, as shown in \cite{Kazantzidis:2018rnb}, in practice an approximated approach is employed which relies on the introduction of correction factors. In our case, and in order to check the results of \cite{Nesseris:2017vor} and \cite{Kazantzidis:2018rnb}, we have chosen the same correction factors used in those references. The same factors were used in \cite{Macaulay:2013swa}. In any case, different approaches can change the significance of the tension.

\section{Testing Planck cosmology}
In order to confront the predictions of standard $\Lambda$CDM model with $f\sigma_8(z)$
measurements, we will obtain theoretical predictions for a general $w$CDM model 
with three free parameters $(\Omega_m, w,\sigma_8)$ with $\sigma_8=\sigma_8(z=0)$. Our benchmark models will correspond to the Planck 2015 and Planck 2018 (TT,TE,EE+lowE) best fit parameters in Table \ref{models}.

\begin{table}
\centering
	\begin{tabular}{c |c| c }
	\hline 
		 & Planck 2015 & Planck 2018   \\ \hline \hline
		 & & \\
		$\Omega_m$ & $0.3156 \pm 0.0091$ & $0.3166 \pm 0.0084$  \\
		$w$ & $-1$ & $-1$  \\
		$\sigma_8$ & $0.831 \pm 0.013$ &$0.8120 \pm 0.0073$  
		 \\ \hline
	\end{tabular}
	\caption{Planck 2015 \cite{Ade:2015xua} and 2018 \cite{Aghanim:2018eyx} (TT,TE,EE+lowE) best fit $\Lambda$CDM parameters.} \label{models}
\end{table}

On the other hand, our data points will correspond to measurements of SDSS \cite{Samushia:2011cs,Howlett:2014opa,Feix:2015dla}; 6dFGS \cite{Huterer:2016uyq}; IRAS \cite{Hudson:2012gt,Turnbull:2011ty}; $2$MASS \cite{Davis:2010sw,Hudson:2012gt}; 2dFGRS \cite{Song:2008qt}, GAMA \cite{Blake:2013nif}, BOSS \cite{Sanchez:2013tga}, WiggleZ \cite{Blake:2012pj}, Vipers \cite{Pezzotta:2016gbo}, FastSound \cite{Okumura:2015lvp} and BOSS Q \cite{Zarrouk:2018vwy}. 
In Table \ref{Datos} we show the 17 independent data points  with the corresponding fiducial 
cosmology parameters corresponding to the so called Gold-2017 compilation of \cite{Nesseris:2017vor} which contains 18 robust and independent measurements based on galaxy or SNIa observations 
together with an additional independent BOSS quasar point. 
As mentioned before, we have removed the two SDSS-LRG-200 points
since they are
based on almost the same galaxy selection as the BOSS-LOWZ point from two heavily overlapping
footprints with BOSS-LOWZ including fainter galaxies.
On the data provided by these surveys we will apply the fiducial cosmology correction given
by (\ref{ratio}).
\begin{table}
\centering
	\begin{tabular}{c c c c c}
	\hline
		Index & Dataset & $z$ & $f\sigma _8 (z)$  & $\Omega _{m}$ \\ \hline \hline
		$1$ & 6dFGS+SnIa & $0.02$ & $0.428 \pm 0.0465$ & $0.3$ \\
		$2$ & SnIa+IRAS & $0.02$ & $0.398 \pm 0.065$ & $0.3$ \\
		$3$ & 2MASS &$0.02$ & $0.314 \pm 0.048$ & $0.266$ \\
		$4$ & SDSS--veloc & $0.10$ & $0.370 \pm 0.130$ & $0.3$ \\
		$5$ & SDSS-MGS & $0.15$ & $0.490 \pm 0.145$ & $0.31$ \\
		$6$ & 2dFGRS & $0.17$ & $0.510 \pm 0.060$ & $0.3$ \\
		$7$ & GAMA & $0.18$ & $0.360 \pm 0.090$ & $0.27$ \\
		$8$ & GAMA & $0.38$ & $0.440 \pm 0.060$ & $0.27$ \\
		$9$ & BOSS--LOWZ & $0.32$ & $0.384 \pm 0.095$ & $0.274$ \\
		$10$ & SDSS-CMASS & $0.59$ & $0.488 \pm 0.060$ & $0.307115$ \\
		$11$ & WiggleZ & $0.44$ & $0.413 \pm 0.080$ & $0.27$ \\
		$12$ & WiggleZ & $0.60$ & $0.390 \pm 0.063$ & $0.27$ \\
		$13$ & WiggleZ & $0.73$ & $0.437 \pm 0.072$ & $0.27$ \\
		$14$ & Vipers PDR--2 & $0.60$ & $0.550 \pm 0.120$ & $0.3$ \\
		$15$ & Vipers PDR--2 & $0.86$ & $0.400 \pm 0.110$ & $0.3$ \\
		$16$ & FastSound & $1.40$ & $0.482 \pm 0.116$ & $0.270$ \\
		$17$ & BOSS-Q & $1.52$ & $0.426 \pm 0.077$ & $0.31$ \\ \hline
	\end{tabular}
	\caption{Data points from \cite{Nesseris:2017vor,Zarrouk:2018vwy}.} \label{Datos}
\end{table}


Apart from the errors quoted in  Table \ref{Datos}, the three points corresponding
to WiggleZ are correlated. Thus the non-diagonal covariance matrix for the 
data points $11, 12, 13$ is given by:

\begin{eqnarray}
   C_{ij} ^{11,12,13} = 10^{-3} \left( \begin{array}{ccc}
 6.4000 & 2.570 & 0.000 \\
2.570 & 3.969 &  2.540 \\
0.000 & 2.540 & 5.184 \end{array} \right)  
\end{eqnarray}
and the total covariance matrix would be
\begin{eqnarray}
C_{ij}  = \left( \begin{array}{cccc}
\sigma _1 ^2 & 0 & 0 & ... \\
0 &C_{ij}^{11,12,13} & 0 & ... \\
0 & 0 & ... & \sigma _N^2
\end{array} \right)
\end{eqnarray}
The corresponding $\chi^2$ is defined as
\begin{eqnarray}
\chi^2(\Omega_m,w,\sigma_8)=V^i C_{ij}^{-1} V^j  \label{chi2eq}
\end{eqnarray}
with $V^i= f\sigma _{8,i} - \textrm{ratio}(z_i) f\sigma _8 (z_i; \Omega_m,w,\sigma_8)$. 
Here $f\sigma _{8,i}$ corresponds to each of the data points in Table \ref{Datos} and  $f\sigma _8 (z_i; \Omega_m,w,\sigma_8)$ is the theoretical value for a given set of parameters values. In order to obtain the two-dimensional confidence regions for the different pairs of parameters, we will construct the marginalized likelihoods integrating the remaining parameter with a flat prior \footnote{The Mathematica code used for the numerical analysis presented in this work is
available upon request from the authors. }, i.e.
\begin{eqnarray}
L(w,\sigma_8)=N\int_{\Delta\Omega_m} e^{-\frac{1}{2}\chi^2(\Omega_m,w,\sigma_8)}\,d\Omega_m
\label{lk}
\end{eqnarray}

 In particular for $\Omega_{m} \in [0.05;0.9]$, $w\in[-2.5;0.5]$ and $\sigma_8\in [0.1;4.0]$. We have checked that the confidence regions remain practically unchanged if we enlarge these intervals.
Notice that this is one of the main differences with respect to 
\cite{Nesseris:2017vor} in which the remaining parameter was fixed to the Planck cosmology value. This procedure implies the introduction of a strong prior in the likelihood (\ref{lk}) from CMB data. However, if we want to determine the confidence regions obtained from  $f\sigma_8$ data alone, no CMB information should be included in  the corresponding likelihoods which is the approach considered in this work.

\section{Results}

\begin{figure}
	\centering
	\includegraphics[scale=0.68]{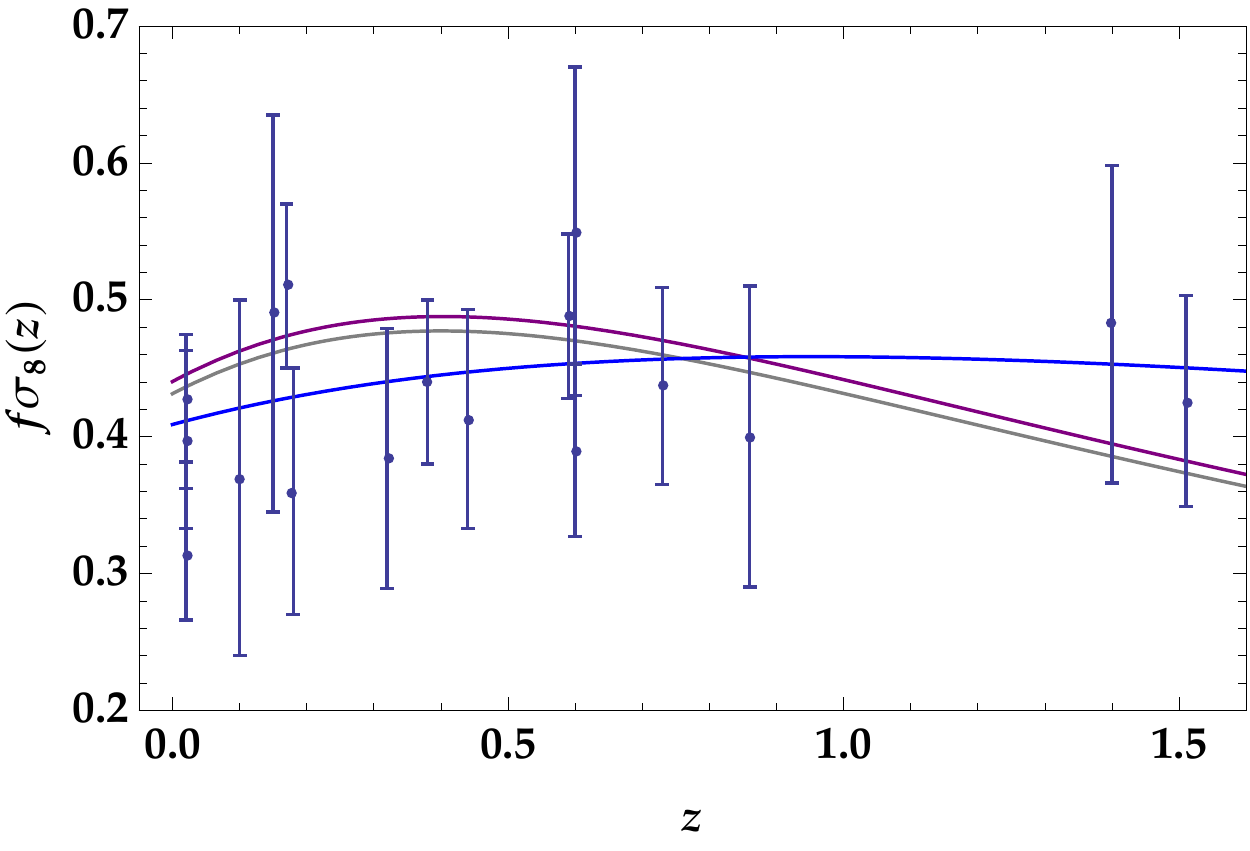}
	\caption{Blue line corresponds to $w$CDM best fit, purple line corresponds to Planck $2015$ and grey line to Planck $2018$.} \label{fig:fsigma}
\end{figure}

In Fig. \ref{fig:fsigma}, the data points quoted in Table \ref{Datos} together with the corresponding $w$CDM best fit curve are represented. The best 
fit corresponds to the parameters $\Omega_{m}=0.145$, $\sigma_8=1.18$ and $w=-0.46$. For the sake of comparison we also show the $f\sigma _8 (z)$ curves corresponding to the Planck $2015$ and Planck $2018$ (in Table \ref{models}) cosmologies. The $\chi ^2$ values for the different models can be found in Table \ref{chi2} together with the corresponding tension level obtained from the $\chi^2$
difference for a three-parameter distribution. As we can see, for both models the tension of Planck cosmology with respect to the best fit $w$CDM cosmology is below 2$\sigma$.

\begin{table}
\centering
	\begin{tabular}{c|c|c}
		\hline
		Model & $\chi ^2_{LRG}$  & $\chi ^2_N$ \\ \hline \hline
		$w$CDM & 16.51 & $10.09$  \\
		Planck $2015$ &21.58 ($1.38\sigma$) & $16.22$ ($1.62\sigma$)  \\
		Planck $2018$ & 18.32 ($0.51\sigma$) & $14.31$ ($1.18\sigma$) \\ \hline
	\end{tabular}
	\caption{Values of $\chi ^2$ for the models plotted in Fig. \ref{fig:fsigma}. 
$\chi^2_{LRG}$ refers to the dataset in \cite{Nesseris:2017vor} whereas 	
	$\chi^2_N$ refers to the data set in this work in Table \ref{Datos}.  The best fit parameters for $w$CDM with $LRG$ data points are $\sigma_8=1.57$, $\Omega_M=0.1$ and  $w=-0.27$. The tension between Planck cosmologies and $w$CDM for three fitted parameters is also shown in parenthesis.} 
\label{chi2}
\end{table}

We see that Planck 2018 provides a better fit than the Planck 2015 cosmology, mainly thanks to the reduction in the $\sigma_8$ parameter, but still both are well above the best fit to $w$CDM.

\begin{figure*}
\begin{center}
	\includegraphics[width=0.45\textwidth]{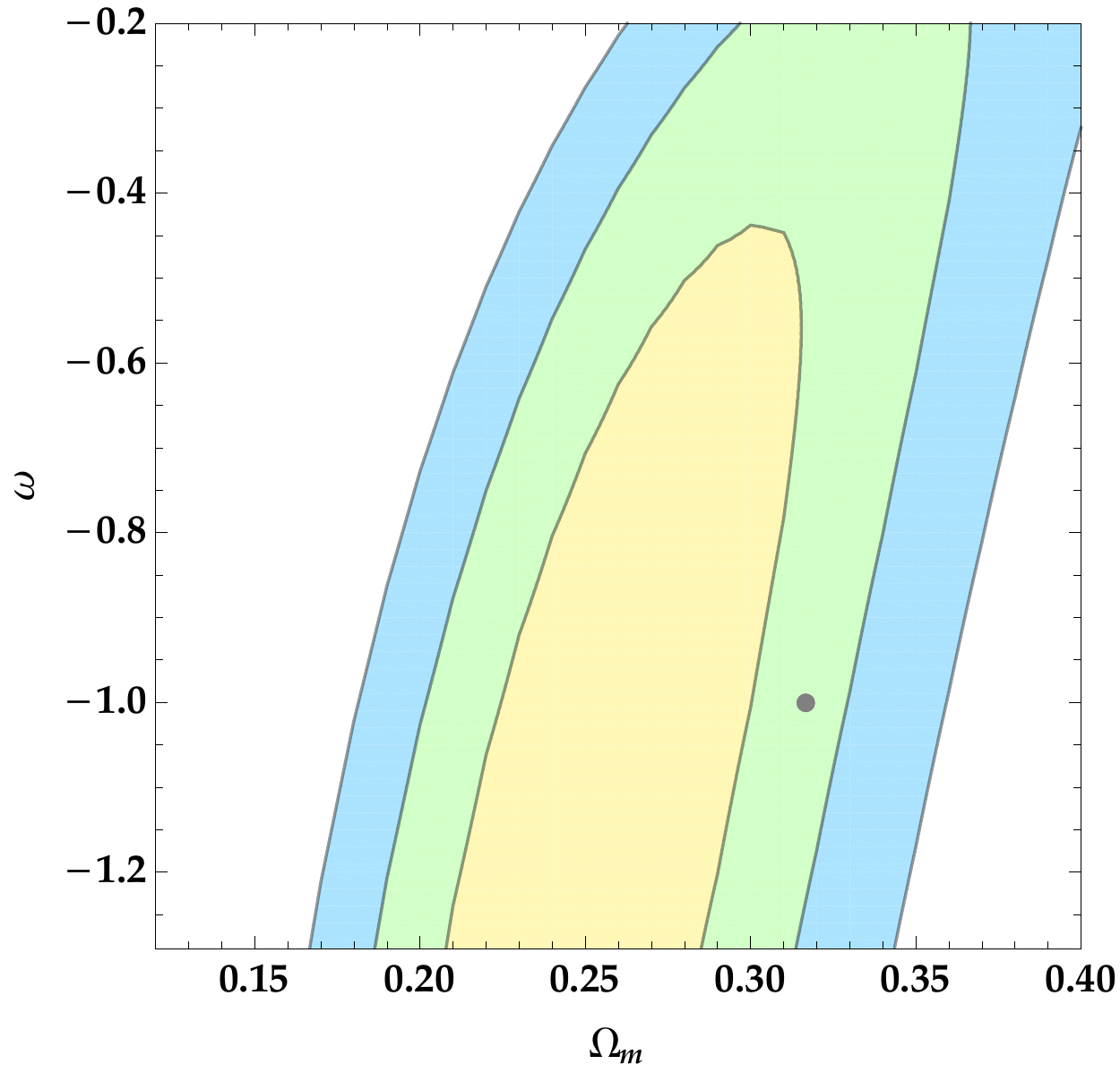}
	\includegraphics[width=0.45\textwidth]{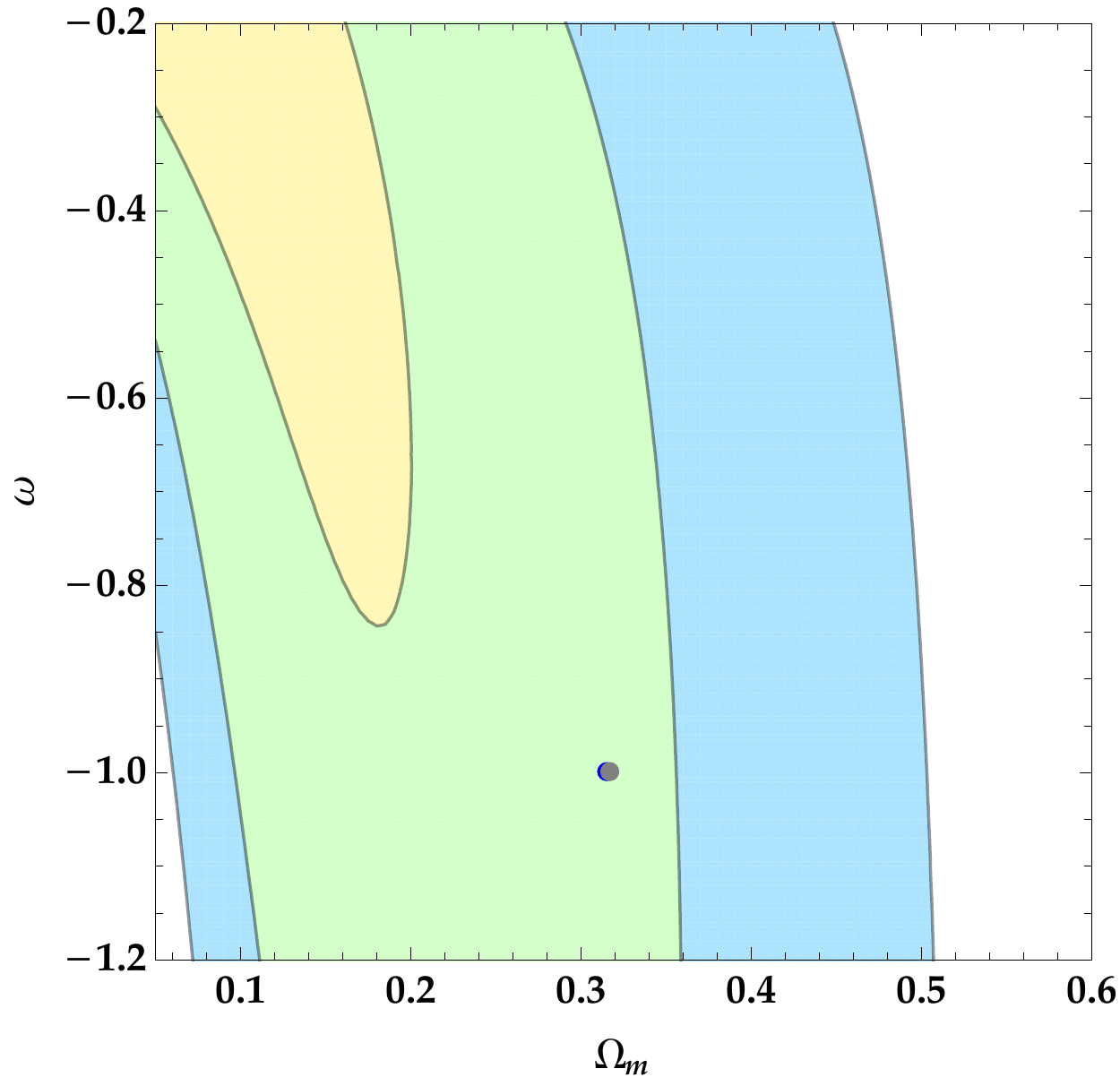}
	\end{center}
	\caption{$w$ vs. $\Omega _{m}$ 1$\sigma$, 2$\sigma$ and 3$\sigma$ confidence regions. Left: maximized contours with $\sigma_8=0.812$. The Planck $2018$ point lays at $1.56\sigma$. Right: marginalized contours. The blue point corresponds to Planck $2015$ and lays at $1.71\sigma$; and the grey point corresponds to Planck $2018$ and lays  also at $1.71\sigma$ } \label{fig:1}
\end{figure*}

\begin{figure*}
\begin{center}
	\includegraphics[width=0.45\textwidth]{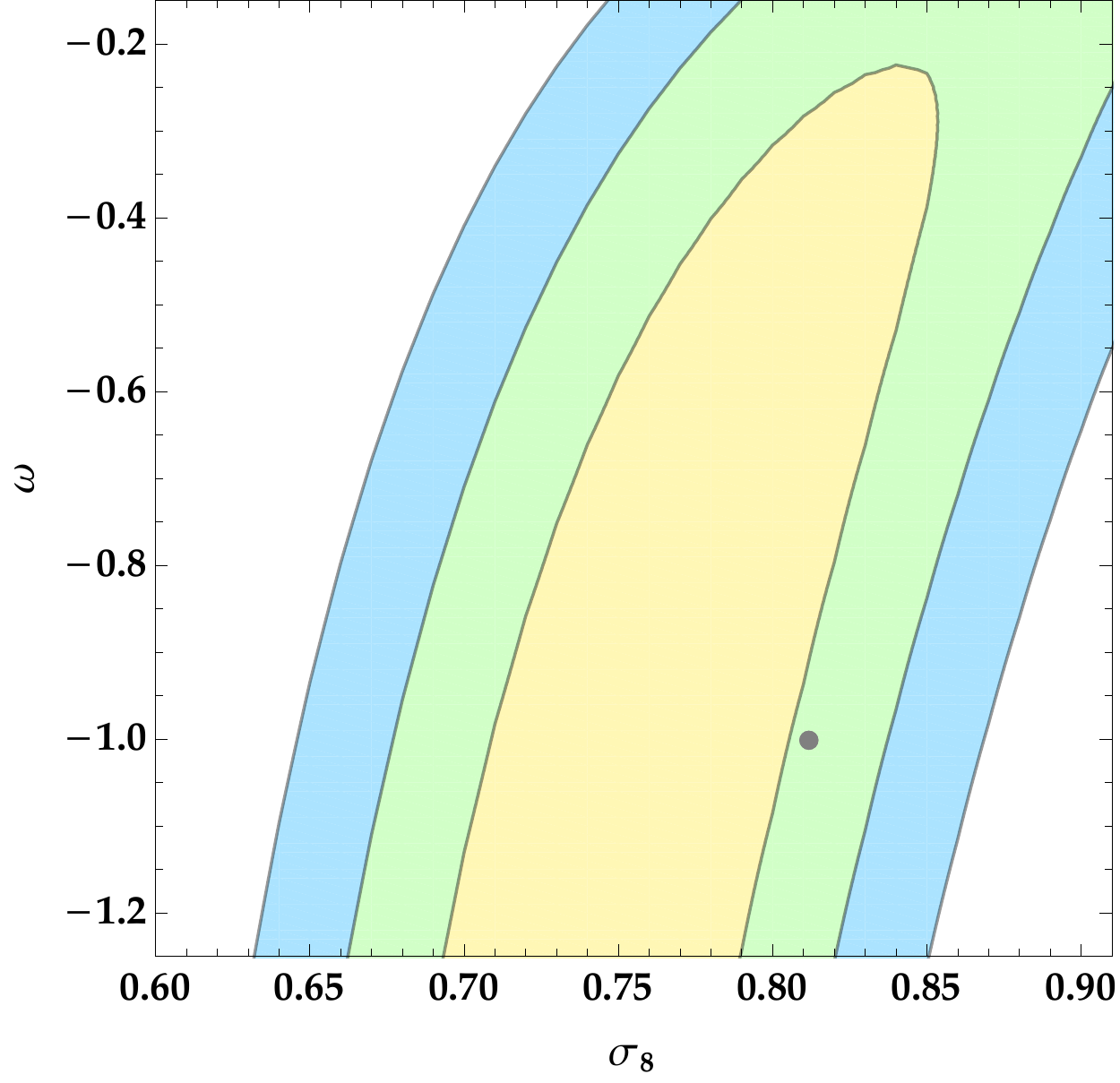}
	\includegraphics[width=0.45\textwidth]{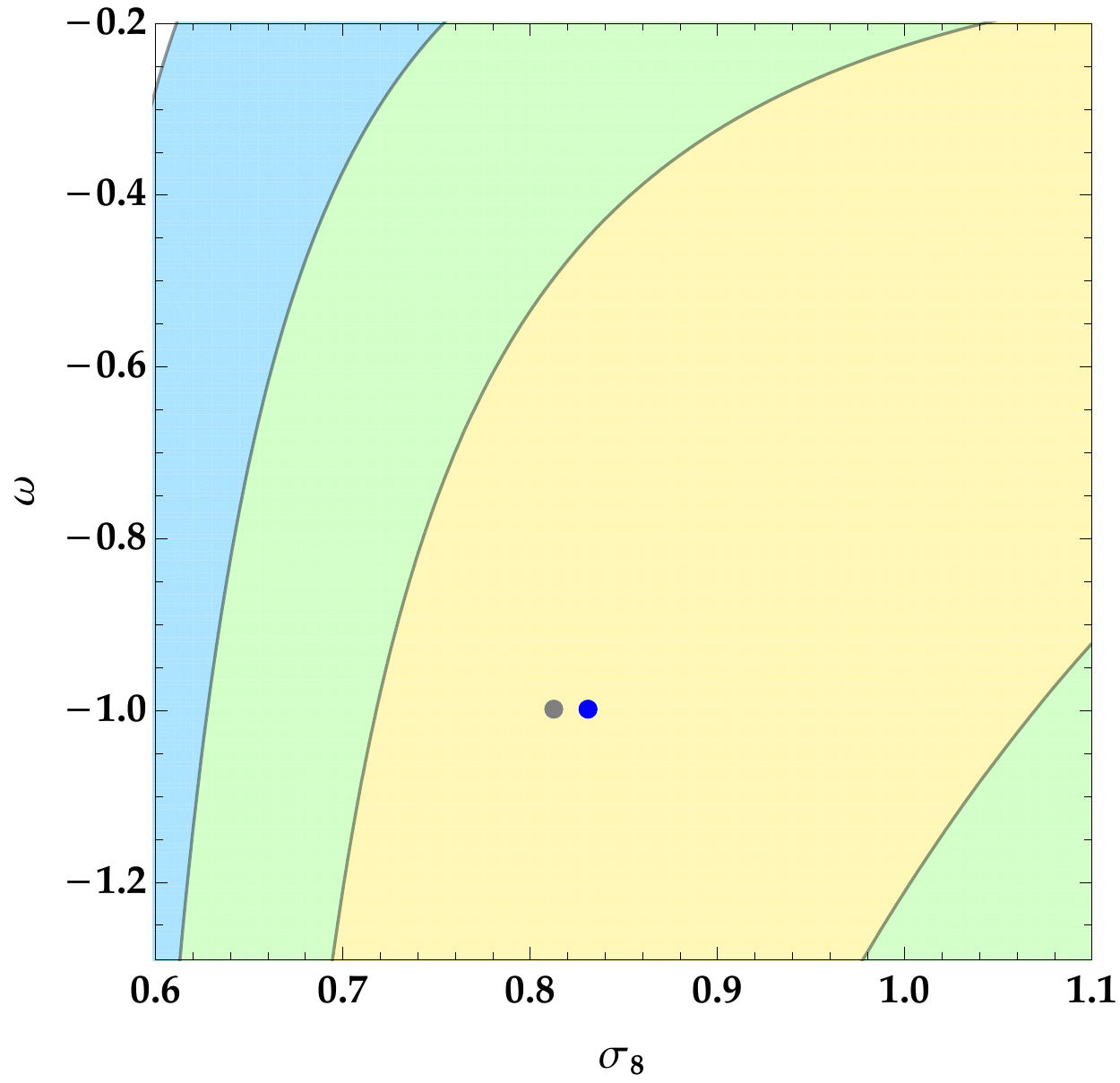}
	\end{center}
	\caption{$w$ vs. $\sigma_8$ 1$\sigma$, 2$\sigma$ and 3$\sigma$ confidence regions. Left: maximized contours with $\Omega_m=0.3166$. The Planck $2018$ point lays at $1.19\sigma$. Right: marginalized contours. The blue point corresponds to Planck $2015$ and lays at $0.26\sigma$  and the grey point corresponds to Planck $2018$ and lays at $0.33\sigma$. } 
	 \label{fig:2}
\end{figure*}

\begin{figure*}
\begin{center}
	\includegraphics[width=0.45\textwidth]{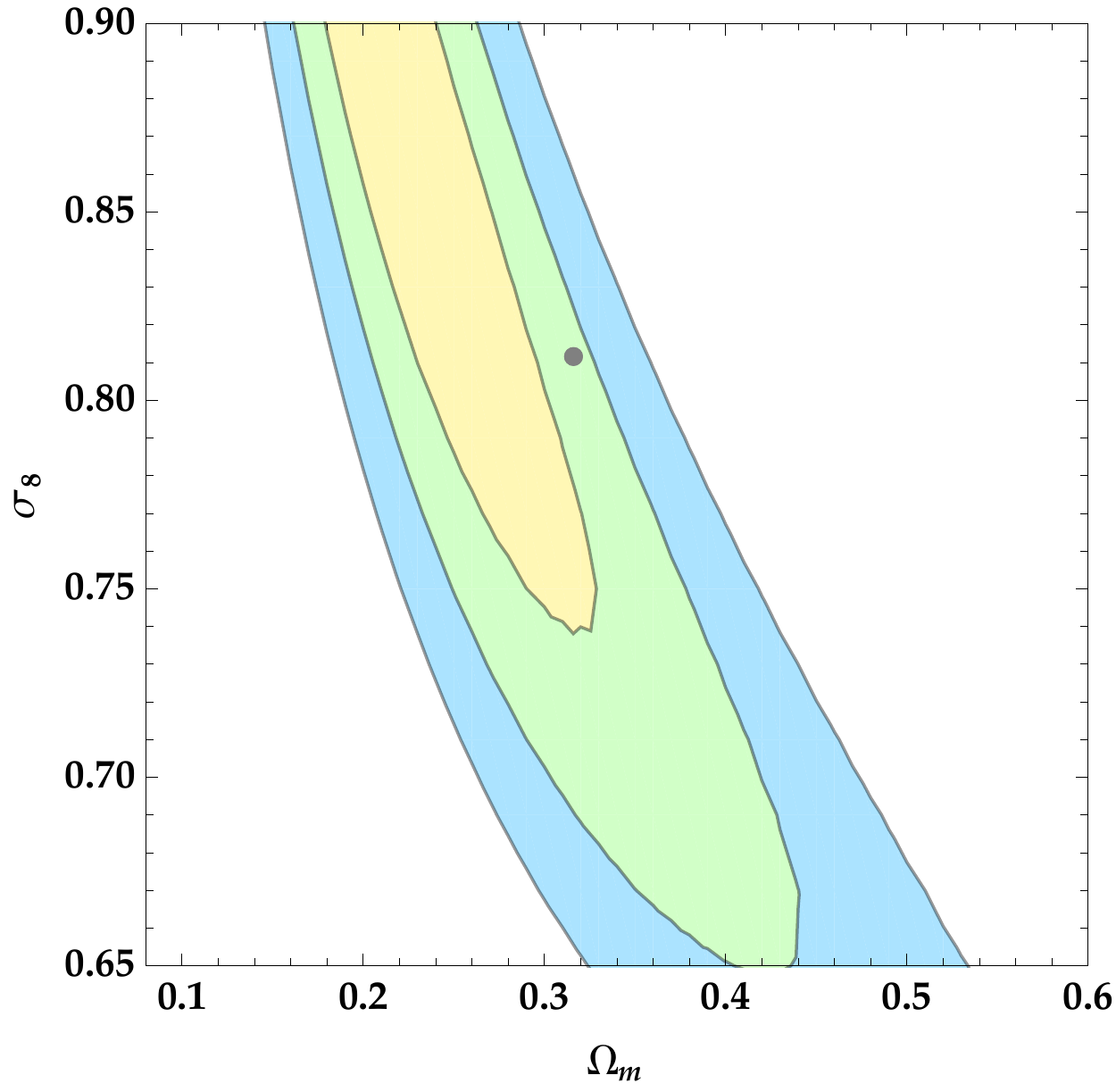}
	\includegraphics[width=0.45\textwidth]{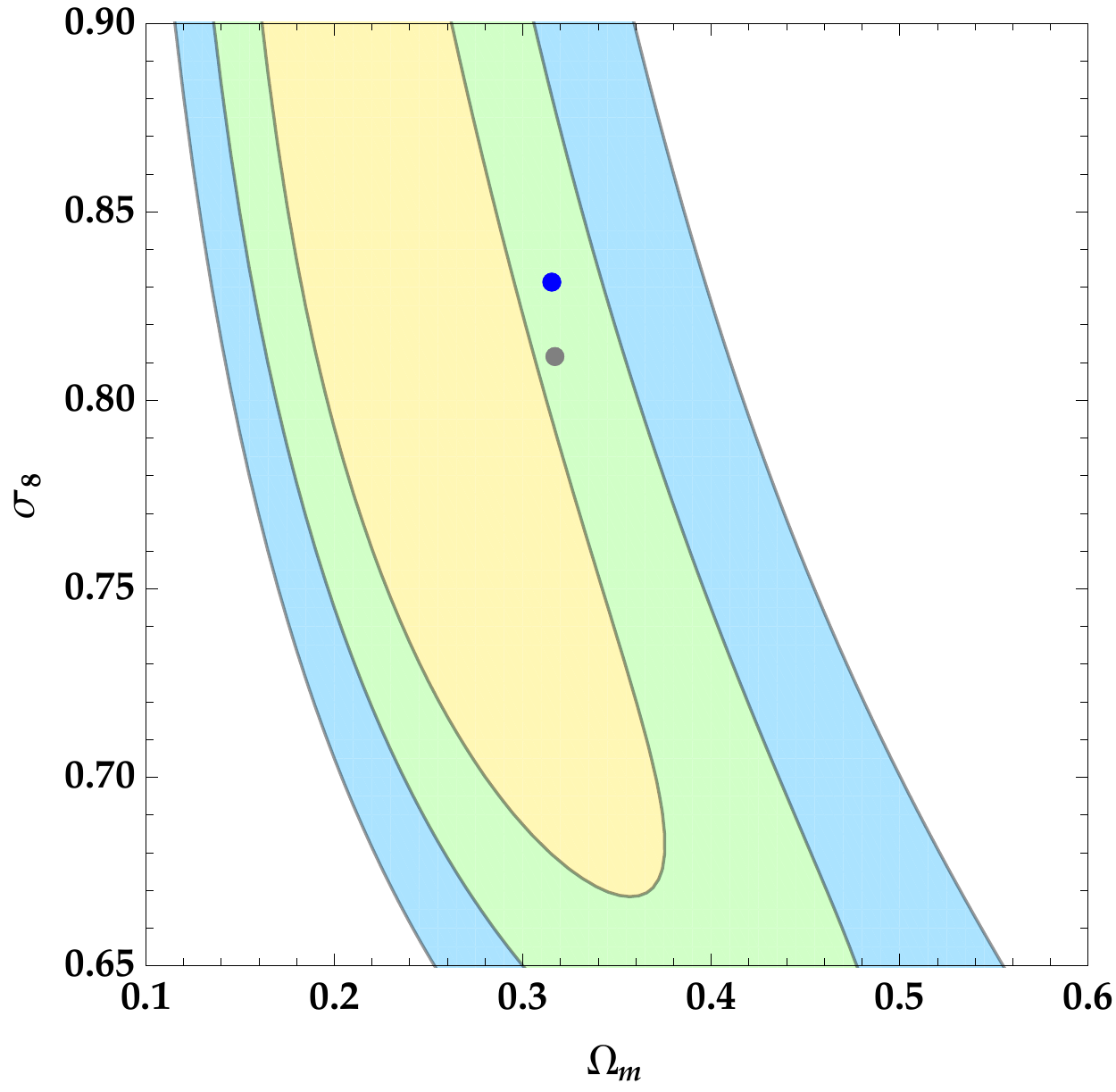}
	\end{center}
\caption{$\sigma_8$ vs. $\Omega_m$ 1$\sigma$, 2$\sigma$ and 3$\sigma$ confidence regions. Left: maximized contours with $w=-1$. The Planck $2018$ point lays at $1.68\sigma$. Right: marginalized contours. The blue point corresponds to Planck $2015$ and lays at $1.44\sigma$. The grey point corresponds to Planck $2018$ and lays at $1.23\sigma$. } 	
 \label{fig:3}	
\end{figure*}

In order to obtain the corresponding confidence regions we will compare two procedures. On one hand, we will follow the approach in \cite{Nesseris:2017vor}
in which the likelihood is maximized, i.e.  in 
the two-dimensional confidence regions the remaining parameter is fixed to the corresponding Planck value
in Table \ref{models}. In the second procedure, the 
remaining parameter is marginalized as mentioned in
the previous section.  In Figs. \ref{fig:1}, \ref{fig:2} and \ref{fig:3} we show the different two-dimensional confidence contours. As we can see, Planck 2018 $\Lambda$CDM shows  tensions of $1.56\sigma$, $1.19\sigma$ and $1.68\sigma$
in the maximized contours which is around $1\sigma$ below the tension found in \cite{Nesseris:2017vor} with
Planck 2015 parameters, partially thanks to the reduced $\sigma_8$ value of Planck 2018 as mentioned before and the exclusion of the two SDSS-LRG points.
On the other hand, the marginalized contours are as expected enlarged as compared to the maximized ones. 
Notice also that although the form of the $(\sigma_8,w)$ and $(\Omega_m,\sigma_8)$ contours are similar in both cases, the marginalization procedure changes the 
shape of the $(\Omega_m,w)$ regions and the tensions with respect to Planck 2018 are   
$1.71\sigma$ for $(\Omega_m,w)$, $0.33\sigma$ for $(\sigma_8,w)$ and
 in the $(\Omega_m,\sigma_8)$ plane we get   $1.23\sigma$.  
In Table \ref{comp} the different tension levels are summarized for the two Planck cosmologies, comparing maximized and marginalized contours and with the dataset in this work and that in \cite{Nesseris:2017vor}.

\begin{table}[H]
		\centering
		\begin{tabular}{c|c|c|c|c|c}
			 \multicolumn{2}{c|}{} & Max$_{LRG}$ & Marg$_{LRG}$ & Max$_{N}$ & Marg$_{N}$  \\ \hline \hline
			\multirow{2}{*}{$\sigma _8$-$w$} & P15 & $2.70\sigma$ & $0.29\sigma$ & $1.76\sigma$ & $0.26 \sigma$ \\ 
			& P18 & $2.04\sigma$ & $0.30\sigma$ & $1.19\sigma$ & $0.33\sigma$ \\ \hline
			 \multirow{2}{*}{$\Omega _m$-$w$} & P15 & $2.86\sigma$ & $1.57\sigma$ & $2.08\sigma$ & $1.71\sigma$ \\
			 & P18 & $2.22\sigma$ & $1.58\sigma$ & $1.56\sigma$ & $1.71\sigma$ \\ \hline
			\multirow{2}{*}{$\Omega_m$ -$\sigma_8$} & P15 & $2.87\sigma$ & $1.50\sigma$ & $2.14\sigma$ & $1.44\sigma$ \\
			& P18 & $2.25\sigma$ & $1.28\sigma$ & $1.68\sigma$ & $1.23\sigma$ \\ 
		\end{tabular}\label{comp}\caption{Tension levels for the Planck 2015 and Planck 2018 $\Lambda$CDM cosmologies in the different two-dimensional maximized and marginalized regions for the datasets in Table \ref{Datos}  ($N$) and in \cite{Nesseris:2017vor} (LRG).}
	\end{table}

\section{Conclusions}

We have revisited the tension of $\Lambda$CDM Planck cosmology with  RSD growth data. We have considered the
Gold data set of \cite{Nesseris:2017vor} together with one additional BOSS-Q point and removing the two SDSS-LRG points thus obtaining a total of 17 independent data points.
   
Confronting these data with the growth rate obtained from a $w$CDM cosmology
with three independent parameters $(w,\Omega_m,\sigma_8)$, we find that unlike previous claims, the tension with Planck 2018 cosmology is below the $2\sigma$ level in all the two-dimensional marginalized confidence regions.
This reduction, which is around $1.5\sigma$ as compared to  \cite{Nesseris:2017vor}, is due to three different factors, namely,  the use of Planck 2018 parameters, the fact that marginalized confidence regions have been considered and the exclusion of the possibly correlated SDSS-LRG points.  Notice that for the $\Lambda$CDM  model (i.e. fixing $w=-1$),  the tension is found to be at $1.68\sigma$ level.

Future galaxy surveys such as J-PAS \cite{Benitez:2014ibt}, DESI \cite{Aghamousa:2016zmz} or Euclid \cite{Laureijs:2011gra} with increased effective  volumes will be able to reduce the error bars in the determination of $f\sigma_8(z)$ in almost an order of magnitude and will help to confirm or exclude the tension analyzed in this work.


\section*{Acknowledgments}

This work has been partially supported by MINECO grant FIS2016-78859-P(AEI/FEDER, UE) and by  Red Consolider MultiDark FPA2017-90566-REDC.




\end{document}